# Spiral Chain $O_4$ Form of Dense Oxygen

Li Zhu[a], Ziwei Wang[a], Yanchao Wang[a], Yanming Ma[a,*], Guangtian Zou[a], and Ho-kwang Mao[b,*]

[a]State Key Laboratory of Superhard Materials, Jilin University, Changchun 130012, China; and [b]Geophysical Laboratory, Carnegie Institution of Washington, Washington, DC 20015

**Oxygen is in many ways a unique element: the only known diatomic molecular magnet and the stabilization of the hitherto unexpected $O_8$ cluster structure in its solid form at high pressure. Molecular dissociations upon compression as one of the fundamental problems were reported for other diatomic solids (e.g., $H_2$, $I_2$, $Br_2$, and $N_2$), but it remains elusive for molecular oxygen, making oxygen an intractable system. We here report the direct prediction on the dissociation of molecular oxygen into a polymeric spiral chain $O_4$ structure (space group $I4_1/acd$, θ-$O_4$) under terapascal (TPa) pressure by using first-principles method on crystal structure prediction. The θ-$O_4$ stabilizes at above 1.92 TPa and has been earlier observed as the high pressure phase III of sulfur. We find that the molecular $O_8$ phase remains extremely stable in a large pressure range of 0.008 – 1.92 TPa, whose breakdown is driven by the pressure-induced softening of a transverse acoustic phonon mode at the zone boundary V point, leading to the ultimate formation of θ-$O_4$. Stabilization of θ-$O_4$ turns oxygen from a superconductor into an insulator by opening up a wide band gap (>3.2 eV) originating from the $sp^3$-like hybridized orbitals of oxygen and the localization of valence electrons.**

# Introduction

As a long-standing problem in the fields of physics, chemistry, and earth and planetary sciences, high pressure dissociation of molecular solids has attracted a lot of attentions, especially on diatomic molecules, such as $H_2$, $N_2$, $O_2$, $F_2$, $Cl_2$, $Br_2$ and $I_2$. Among these molecular systems, solid oxygen is of particular interests and exhibits many unusual physical properties by virtue of its molecular spin and the resultant spin-spin interactions, which affect its physical properties and make the system a critical test of condensed-matter theory (1, 2). Oxygen is also the third most abundant element in the solar system, and its behavior under extreme pressures provides important insight into the oxygen-related systems, which is of crucial importance for the understanding of the physics and chemistry of planetary interiors.

Oxygen exhibits a rich polymorphism with seven crystallographic phases unambiguously established. On cooling at ambient pressure, oxygen is in turn solidified to the paramagnetic γ phase, magnetically disordered (short-range ordered) β phase (3, 4), and eventually antiferromagnetic α phase (5). At ~6 GPa, the α phase transforms into the antiferromagnetic δ-phase (6-8). Under a higher pressure of ~8 GPa, the magnetic order of oxygen was destroyed at the formation of the ε-phase consisting of $O_8$ clusters (9, 10). The ε-$O_8$ phase displays the bonding characteristics of a closed-shell system, where the intermolecular interactions primarily involve in the half-filled 1 $\pi_g$* orbital of $O_2$ (11). Above 96 GPa, ε-$O_8$ was observed to transform into a metallic ζ-phase (12, 13), showing an intriguing superconductivity with a transition temperature of 0.6 K (14). The superconducting ζ-phase was later on theoretically predicted (15) and then experimentally (16) confirmed to adopt the ζ-$O_8$ structure, which is isostructural with ε-$O_8$ but with a shorter nearest O-O distance between $O_8$ clusters than the inter-$O_2$ distance within the $O_8$ cluster[23]. Although all these studies shed light towards a complete understanding of the high-pressure phase diagrams of solid oxygen, the report on the formation of polymeric structure up to the highest pressure of 500 GPa (15) studied is scarce, below which other diatomic molecules (17-21) show readily molecular dissociations.

Here, we have extensively explored the high-pressure crystal structures of solid oxygen up to 2 TPa using the first-principles method on crystal structure prediction (22), which was designed to conduct global minimization of free energy surfaces merging *ab initio* total-energy calculations via CALYPSO (Crystal structure AnaLYsis by Particle Swarm Optimization) methodology (22). Our CALYPSO method has been implemented in the CALYPSO code (23) and successfully benchmarked on various known systems (22). Several predictions on high pressure structures of dense Li, Mg, Bi$_2$Te$_3$, and water ice (24-27), were successfully made, where the predicted high pressure insulating *Aba*2-40 (Pearson symbol oC40) structure of Li and the two low-pressure monoclinic structures of Bi$_2$Te$_3$ were confirmed by independent experiments (26, 28).

## Results and Discussion

Our structure searches through the CALYPSO code with system sizes up to 24 atoms per simulation cell were performed at pressures of 0.02 - 2 TPa. The structural simulations at 0.02 TPa were able to successfully predict the experimental ε-O$_8$ structure, while at 0.1-1.5 TPa, ζ-O$_8$ is predicted to be most stable, in agreement with earlier prediction (15) and experimental observation (16). At 2 TPa, we surprisingly discovered an energetically most stable tetragonal *I*4$_1$/*acd* structure (named as θ-O$_4$, 16 atoms /cell, Fig. 1a), which is ~1.3% slightly more denser than ζ-O$_8$. Remarkably, the molecular feature in θ-O$_4$ structure is no longer preserved; instead, squared chains with four atoms per turn are formed. There, each oxygen has two identical nearest neighbors with the nearest O-O distance of 1.153 Å at 2 TPa, much larger than the intra-molecular distances (e.g., 1.036 Å at 1.8 TPa) in ζ-O$_8$. The chains are formed along the *c*-axis, with an O-O-O bond angle of 98.79°. The nearest-neighboring O-O distance between chains is 1.552 Å, which is ~35% longer than the intra-chain O-O distance. The calculated enthalpy curve (Fig. 2a) at a higher level of accuracy for θ-O$_4$ confirms its energetic stability relative to ζ-O$_8$. Phonon calculations (supplementary

Fig. S2) established the dynamical stability of the θ-O$_4$ structure since we do not find any imaginary phonons in the whole Brillouin Zone.

Oxygen is the lightest element in Group VI of the periodic table, with sulfur being the next heavier family member. Previous theoretical studies assumed that the molecular ring structure of S$_8$ (S-I) in sulfur might be one of the candidates for the high-pressure phases of solid oxygen (29). However, energetic calculations suggested that oxygen in the S$_8$ structure is energetically very unfavorable if compared with other known phases of oxygen. Interestingly, we here found that θ-O$_4$ shares the same structure type with the high-pressure S$_4$ phase (30) (S-III) of sulfur. The reasons why atomic oxygen is unable to adopt the lower pressure phases (S-I and the trigonal chain S-II) of sulfur might be stemmed from that oxygen has a very small 1$s$ core (the 1$s$ orbital radius of oxygen is 0.068 Å) capable of forming very short bonds. This small 1$s$ core in oxygen is in apparently contrasted to the much larger 2$p$ core in sulfur (the 2$p$ orbital radius of sulfur is 0.169 Å).

To understand the physically driven force of the formation of this polymeric θ-O$_4$ structure, we calculated the phonon dispersion curves of ζ-O$_2$ with elevated pressures. It is found that a transverse acoustic (TA) phonon mode at the zone boundary *V* point softens with pressure (Fig. 3). We subsequently plotted out the eigenvectors of this TA(*V*) phonon mode (inset of Fig. 3), where the arrows represent the directions of atomic vibrations. It is clearly seen that this softening phonon mode involves in the planar rotation of oxygen molecules around the *a*-axis. The role of this rotation is to effectively shorten the inter-molecular O-O distance and enlarge the intra-molecular bond, which ultimately leads to the stabilization of the squared chains in θ-O$_4$. It is noteworthy that though at the phase transition pressure of 1.92 TPa, the TA(*V*) phonon frequency in ζ-O$_2$ doesn't soften to zero, it can still initiate the phase transition. This transition mechanism shares the similarity with those observed in SiO$_2$ (31) and MgH$_2$ (32), where the rutile→CaCl$_2$ phase transition was driven by the softened non-zero optical mode at zone center (32).

The metallization of ζ-O$_2$ is attributed to the creation of nearly free electrons by the pressure-induced band overlap (15). As a result, a very low electronic density of states at Fermi level is achieved, in a close correlation with the observed low superconducting transition temperature (0.6 K) (14). At the formation of atomic θ-O$_4$ phase, solid oxygen remarkably turns to an insulator as suggested by the simulated band structure and partial electronic density of states at 2 TPa (Fig. 2b). The calculated band gap is ~3.2 eV--it should be much larger as the density functional calculation underestimates significantly the band gap. To probe the physical mechanism of this insulation, we have performed the chemical bonding analysis of θ-O$_4$ through the calculations of electron localization function (ELF) as shown in Fig. 1 (c) and (d), where the calculated ELF isosurface clearly indicates two lone pairs and two O-O covalent bondings for each oxygen atom. This allows us to identify the peculiar $sp^3$–like bonding orbitals of oxygen in θ-O$_4$. As a consequence, a full localization of valence electrons is found, able to create a wide gap insulator. The intriguing bonding behavior of oxygen in θ-O$_4$ is not fully unexpected since it somewhat resembles with the chemicals of oxygen in water ice with the formation of two O-O single bonds. What makes it unique is that the $sp^3$ orbitals of oxygen in θ-O$_4$ are distributed along the infinite chains, unlike the perfect localization in the isolated water molecules. Notably, the O-O-O bond angle is 98.79° in θ-O$_4$, which is smaller than 109° for an ideal tetrahedral angle. This deviation can be naturally explained by a Coulomb repulsive interaction between the lone pairs.

The insulating feature in θ-O$_4$ is in contrast to the metallicity of the high pressure atomic phases of solid H$_2$ (17) and halogens (I$_2$ and Br$_2$) (18, 19). This is originated from the fact that hydrogen and halogen elements having one and seven electrons in their outer shells, respectively, are unable to form the perfect saturation bonding in the atomic structures favorable for the metallicity. This θ-O$_4$ insulation seemingly resembles the polymeric nitrogen (21, 33), where five valence electrons of each N atom form three covalent bonds and a lone pair, resulting in the complete localization of the valence electrons. However，θ-O$_4$ is distinctly created from the compression of a metallic solid, which is clearly against the chemical intuitions, as earlier exemplified by the remarkable findings of metal-insulator transitions at high pressures in lithium (25, 34) and sodium (35). The predicted formation of θ-O$_4$ in solid oxygen represents a significant step forward in understanding the high-pressure phase diagrams of solid oxygen, and might improve our understanding of other oxygen-related materials at extreme conditions, such as the planetary interiors of giant planets.

## Methods

The underlying *ab initio* structural relaxations and electronic calculations were performed in the framework of density functional theory within the Perdew-Burke-Ernzerh (PBE) and, as implemented in the VASP (Vienna *ab Initio* Simulation Package) code (36). The all-electron projector augmented wave (PAW) (37, 38) method was adopted, with the PAW potential treating $1s^2$ as core. The cutoff energy (910 eV) for the expansion of the wave function into plane waves and Monkhorst-Pack *k* (39) meshes were chosen to ensure that all the enthalpy calculations are well converged to better than 1 meV/atom. The phonon calculations were carried out by using a supercell approach (40) as implemented in the PHONOPY code (41). The validity of the used PAW potential for the currently studied high pressures (≤ 2 TPa) was carefully checked by comparing with the full-potential all-electron calculation through WIEN2K code (42). In the full-potential calculations, the muffin-tin radii were chosen to be 0.97 a.u. for O. The plane-wave cutoff was defined by $K_{max}R_{MT} = 7$, where $R_{MT}$ represents the muffin-tin radius and $K_{max}$ the maximum size of the reciprocal-lattice vectors. Convergence test gave the choices of 1000 *k* points for both ζ-O$_8$ and θ-O$_4$ phases in the electronic integration of the Brillouin Zone.

## Acknowledgements

The authors acknowledge the High Performance Computing Center of Jilin University for supercomputer time and funding from the National Natural Science Foundation of China under grant Nos. 11025418 and 91022029, and the China 973 Program under

This is a pre-print version of the following article:
Li Zhu et al, Spiral chain $O_4$ form of dense oxygen, *Proc. Natl. Acad. Sci. U.S.A.* (2011), doi: 10.1073/pnas.1119375109, which has been published online at http://www.pnas.org/content/early/2011/12/27/1119375109

Grant No. 2011CB808204.## Author Contributions

Y. M. proposed the research. L. Z., Z. W., Y. W. and Y. M. did the calculations. L. Z., Z. W., G. Z. Y. M., and H. M. analyzed the data and wrote the paper. L. Z. and Z. W. contributed equally to this work.

*Correspondence and requests for materials should be addressed to Y.M. (mym@jlu.edu.cn) or H. M. (h.mao@gl.ciw.edu)7

## Figure captions

**FIG. 1**. (color online) (a) and (b) The predicted θ-O$_4$ structure viewed along the *c* and *a* axes, respectively. The optimized lattice parameters at 2 TPa are *a* = 5.651 Å and *c* = 1.803 Å with oxygen occupying 16*f* (0.1359, 0.3859, 0.125) positions. (c) The calculated ELF Isosurface of θ-O$_4$ structure with ELF = 0.8. (d) ELF plots in (100) section.

**FIG. 2**. (a) Calculated enthalpy per atom as a function of pressure with respect to ζ-O$_8$. For a double check, inset of (a) shows the enthalpy of θ-O$_4$ relative to ζ-O$_8$ obtained by a full-potential all-electron calculation through WIEN2K code. The all-electron method gives basically the same result on the transition pressure with only a < 2% difference, which confirms the validity of the PAW potential adopted here at such high pressures. (b) Band structure (left panel) and partial electronic density of states (DOS, right panel) of θ-O$_4$ at 2 TPa.

**FIG. 3**. (color online) Phonon frequencies of the softened transverse acoustic phonon mode at the zone boundary *V* point with pressure in ζ-O$_8$. Inset: The phase transition mechanism of ζ-O$_8$ → θ-O$_4$. Arrows indicated the eigenvector of this softened mode.

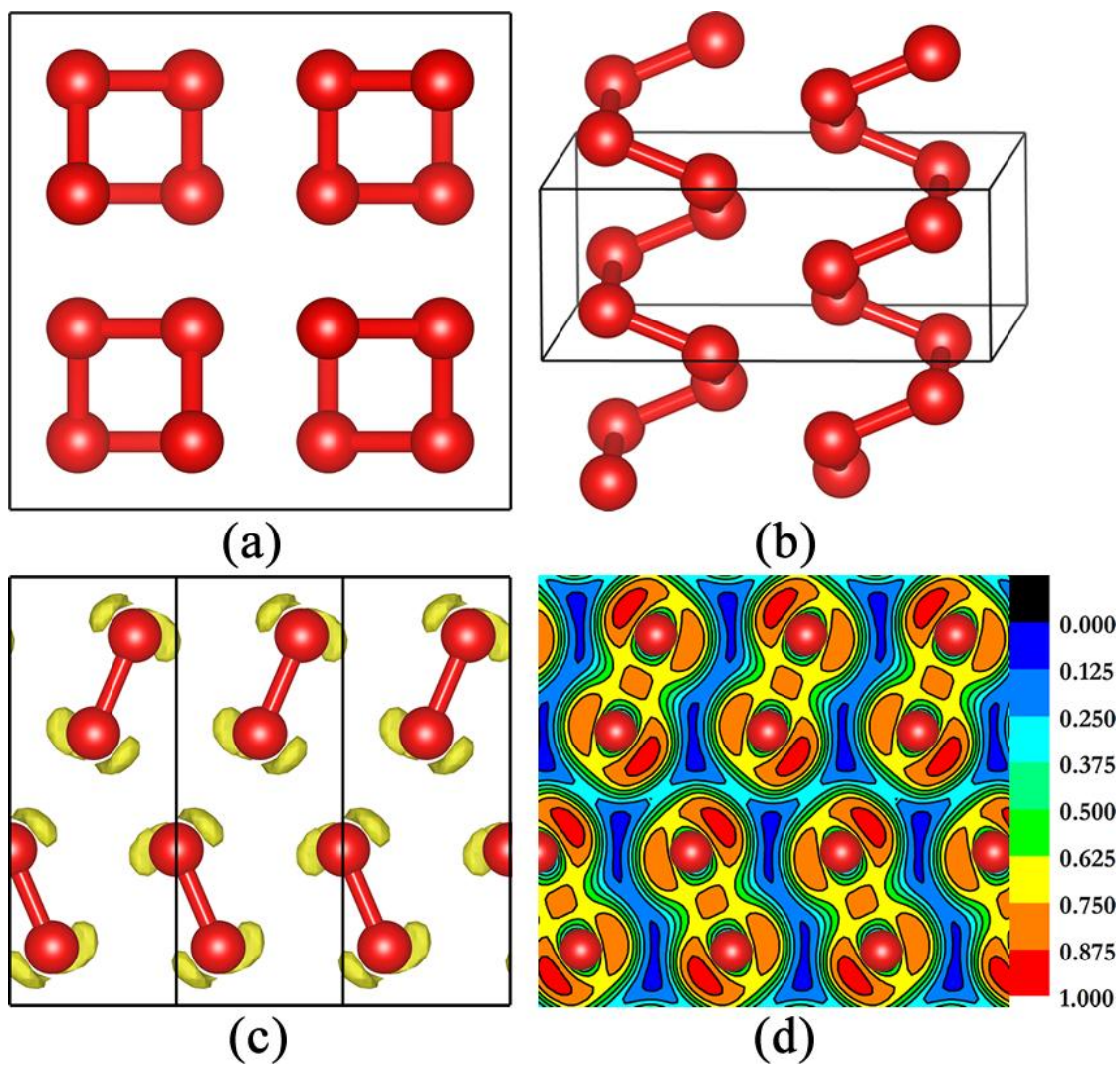

FIG. 1

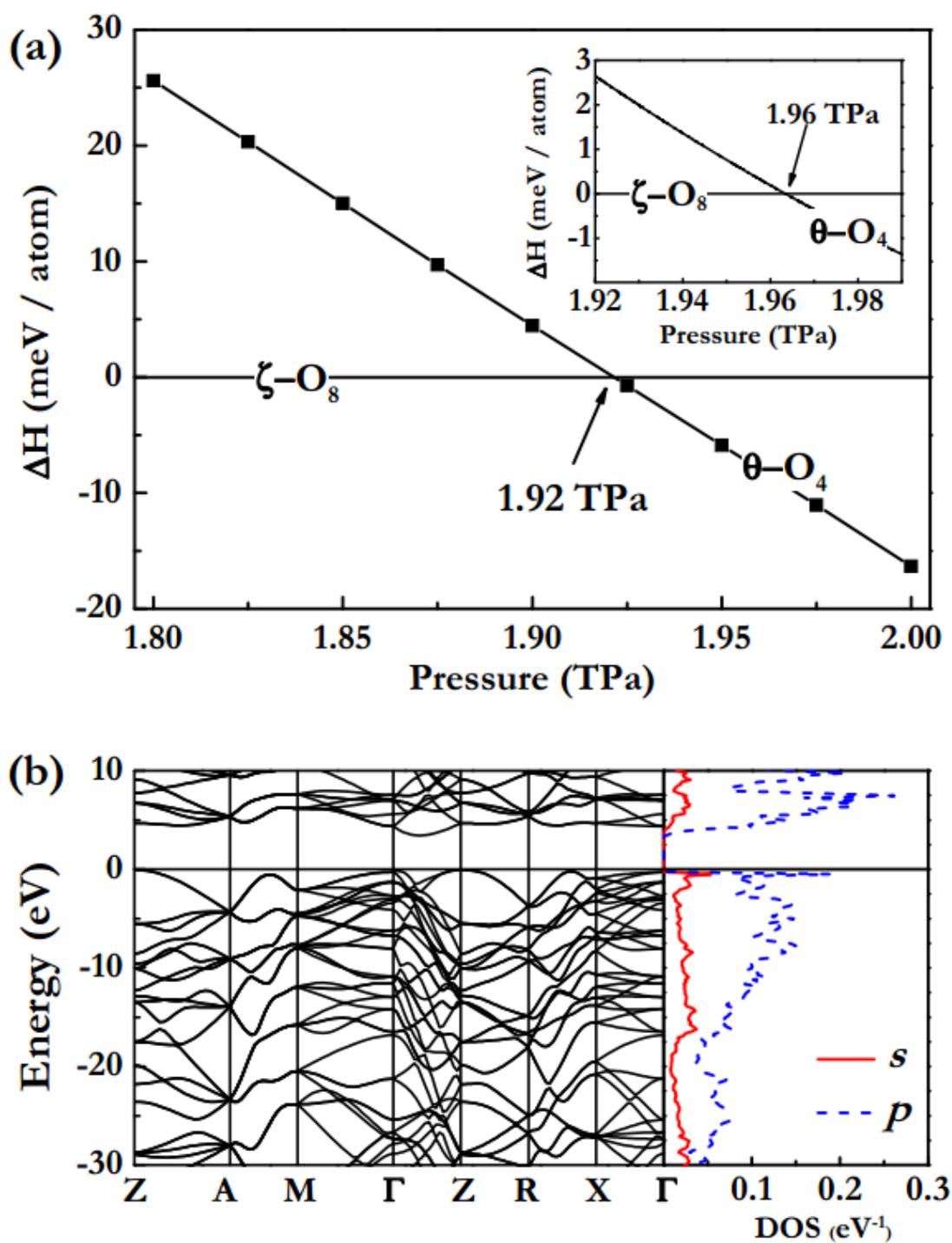

FIG. 2

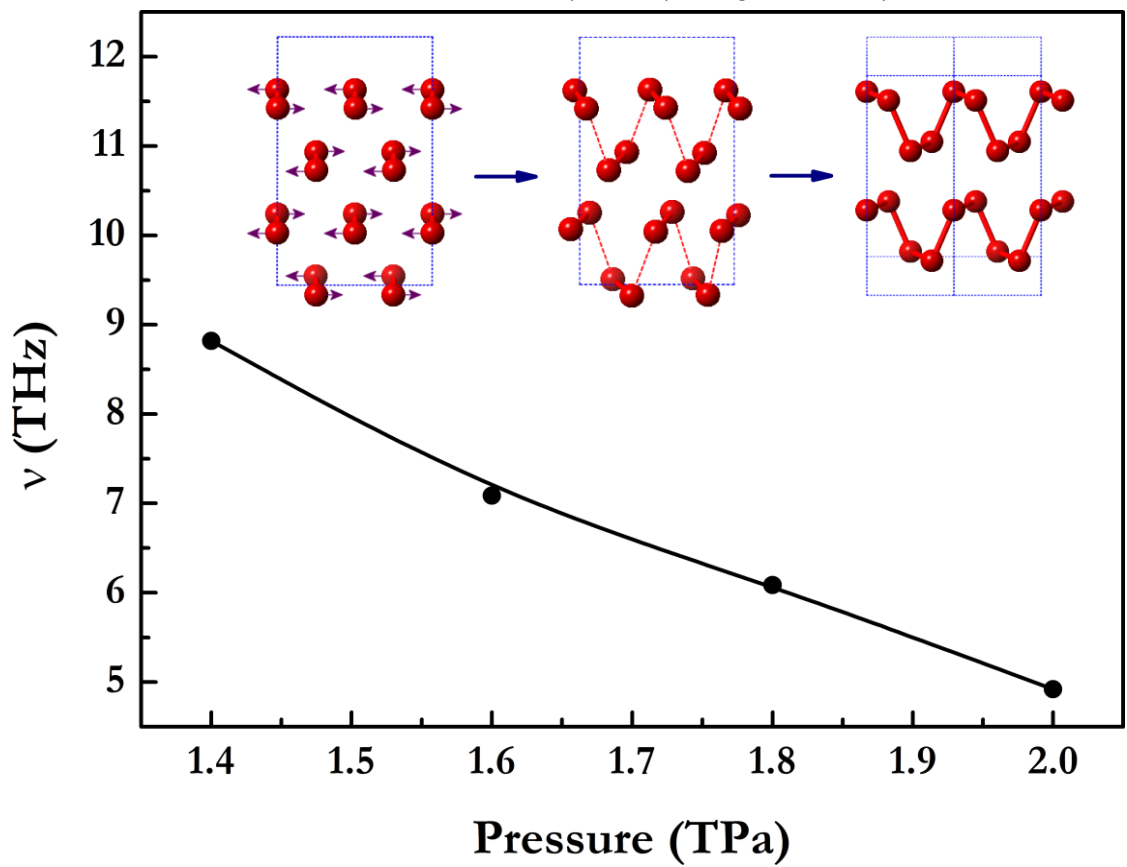

FIG. 3